\documentclass[conference,10pt]{IEEEtran}
\usepackage[utf8]{inputenc}
\usepackage{graphicx}
\usepackage{xspace}

\newcommand{\archname}{EBA\xspace}

\newcommand{\principlename}{DST\xspace}

\title{Exposed Buffer Architecture for Continuum Convergence}

\author{\IEEEauthorblockN{Micah Beck}
\IEEEauthorblockA{University of Tennessee\\
Knoxville, Tennessee 37996\\
Email: mbeck@utk.edu}
\and
\IEEEauthorblockN{Terry Moore}
\IEEEauthorblockA{University of Tennessee\\
Knoxville, Tennessee 37996\\
Email: tmoore@icl.utk.edu}}

\date{July 2020}

\begin{document}

\maketitle
\begin{abstract}
Exposed Buffer Architecture addresses the problem of creating a programmable service platform for the digital continuum by reexamining the particular form of virtualization that is inherent to the Internet architecture. In the Internet stack below the Network Layer, the Link layer models services that are local to network nodes, or that connect them in local area networks. Aggregating these low level resources in the implementation of IP to create wide area services serves two different purposes: 1) It virtualizes local services, enabling interoperability through the adoption of a common model, and 2) It hides the topology of local infrastructure. The main premise of Exposed Buffer Architecture is that we can separate these two tasks, addressing the first with an invariant system model that provides a highly general, programmable platform for transcontinuum services, enabling a variety of approach to address the second.
\end{abstract}

\section{Introduction}
Three trends are combining to produce the impression that we are in the midst a technological sea change. First, there is the explosive growth in the number, diversity, and pervasiveness of digital devices and systems across the entire range of human activity. This emerging landscape has come to be called the ``computing \dots’’ or “digital continuum”~\cite{Panziera2020etp4hpc}. The second trend, closely tied to the first, is the ongoing, exponential increase in the volume of data being generated \emph{outside} of data centers and commercial clouds in the \emph{data periphery}, i.e., in the vast expanse of the digital continuum that extends from the border of the data center to extreme edges of the network. Finally, there is the arrival of revolutionary new techniques for machine learning and AI that automate the improvement of the empirical rules that enable us to control these systems and make them serve our purposes. This confluence of factors has led to a search for a new paradigm for network computing. Below we describe the \emph{Exposed Buffer Architecture} (EBA), which offers a radical new approach to combining storage, networking and computing in a \emph {converged platform} for creating ubiquitous services on the digital continuum.

Computing on memory-resident data and store-and-forward networking both require memory and processor cycles.  Thus a ``services everywhere’’ vision of a smart and data-intensive digital continuum can only be based on a distributed platform that somehow converges computing, networking, and storage. The idea that computing and networking could and would converge has been a constant theme of the Internet era, from Sun Microsystem’s ``the Network is the Computer’’ in the 1980’s, down to the ``converged infrastructure’’ and ``hyper-converged infrastructure’’ that are now staples of marketing to data centers. In his prescient analysis of the convergence of computing and telecommunications, David Messerschmitt pointed to the way in which industries based on networked applications (e.g., telephony, video, data), and which had emerged as vertically integrated ``stovepipes’’ or ``silos’’ built on dedicated infrastructure, were seeing the boundaries between them dissolved. 
Messerschmitt attributed this trend to the broad adoption of common interfaces for layered middleware services — ``horizontal integration’’— that all could build on~\cite{messerschmitt1996convergence}. The most important factor in this momentous ``Internet convergence’’ was the near universal adoption in the 1990’s of the Internet protocol (IP) as a \emph{spanning layer} for networking. 

A spanning layer is a common interface implemented across an infrastructure in order to provide a homogeneous way of using the heterogeneous resources in the layers beneath it, while also facilitating the adoption of standard protocols and interfaces above it~\cite{clark1995interoperation}. Following in this tradition, we believe the main problem in creating a viable platform for service creation and innovation on the digital continuum lies in the design of a successful spanning layer that converges the computing, networking, and storage/memory silos we now take for granted. In other words, building a ``services everywhere'' digital continuum means desiloing cyberspace. 

\archname is based on the observation that networking, processing and storage are higher level services whose implementations are all comprised of operations on persistent buffers (see Figure~\ref{Venn}). But even if, because of this fact, all of these services could be expressed in terms of a common primitive buffer service, why should we try to do so?
Our primary motivation is to define an architecture for a shared infrastructure for the implementation of services on the digital continuum that span those traditional ICT silos and that will be as widely used as possible for as long as possible. According to our analysis, the key to achieving such a vision is a spanning layer for EBA that maximizes ``deployment scalability.''

\iffalse 
In a stack of services higher layers are implemented in terms of lower layer ones.
A "spanning layer" is a layer that virtualizes those below it. Layers above it must be expressed in terms of the spanning layer~\cite{clark1995interoperation}.
The prototypical example of spanning layers are the Internet Protocol Suite at the Network layer of the Internet stack that virtualizes local area services.
Another example of a spanning layer is the Unix/POSIX operating system call interface which virtualizes the architectural resources of the host computer.

Any spanning layer creates a community of interoperability in applications that are expressed at layers above it on the stack.\fi

{\it Deployment scalability} is a term that denotes the tendency of a spanning layer to be widely and willingly adopted in ways that span boundaries (what one might call "viral growth")~\cite{beck2019hourglass}.
Another way of characterizing deployment scalability is that it means that adoption of the spanning layer offers benefits that outweigh those of using a specialized interface.
The \emph{Deployment Scalability Tradeoff} (\principlename) is a principle which tells us that there is a correlation between the deployment scalability of a spanning layer and the degree to which it is logically weak, simple, general and limited in its allocation of resources. In designing EBA, we have used this principle as a tool for analyzing the effect of different design choices on the deployment scalability of the common buffer service at its foundation.
%In the context of EBP, this translates into a minimal and orthogonal API that specifies service at the local layer, which is inherently weaker logically than a globally defined service.
%This motivates EBP being a best effort service which has limits on all atomic resource allocations and requiring aggregation of those resources to meet specific reliability and performance goals.
%
%While deployment scalability is the goal of most infrastructure service definitions, it has proved difficult to achieve in many domains, including programmable and stateful networking.
%The Deployment Scalability Tradeoff is a principle which relates the design of a spanning layer to its deployment scalability.

Another part of the problem is that a successful spanning layer is a double edged sword. On one hand, the community of interoperability it creates will, as its deployment scales up, generate positive network effects and innovations that benefit a huge community users. On the other hand, as the set of adopters grows, the interface becomes progressively more ossified and the costs of choosing alternatives get ever higher. This downside of spanning layer success was highlighted in 2005 by leaders of the networking research community in an impassioned call to “[Overcome] the Internet Impasse through Virtualization”~\cite{anderson2005overcoming}. These researchers noted that the current Internet architecture had become so deeply entrenched that proposed substantial changes (e.g., for security and QoS) were effectively undeployable in real world environments, even for the purposes of experimentation and testing. Consequently, Internet Service Providers were being forced to deploy a diverse ensemble of ``\dots \textit{ad hoc} work-arounds, many of which violate the canonical architecture (e.g., middleboxes) \dots in order to meet legitimate needs that the architecture itself could not. These architectural barnacles \dots may serve a valuable short-term purpose, but significantly impair the long-term flexibility, reliability, and manageability of the Internet.'' Fifteen years later, this ``impasse,'' or state of stagnation in network innovation, remains essentially unchanged~\cite{mccauley2019enabling}.

\section{Two Perspectives on Convergence}

% Network Function Virtualization offers Internet Service Providers (ISPs) a way to address many issues raised by the growing menagerie of middle- and edge-box ``barnacles'' by implementing the functions they provide in software, thereby making it possible to combine them more seamlessly with core network functionality running on commodity hardware. Specifically, NFV creates a space within an IP router to execute processes that implement a variety of functions while maintaining the performance of packet forwarding along the fast (NFV-independent) routing path. The Exposed Buffer Architecture (\archname) described below shares with NFV the general goal of enabling greater network innovation, but takes a more radical approach to the ``Internet impasse,'' one which we believe opens up a path for significant innovations that conform in a much more principled way to the architecture of the underlying platform.

\iffalse
As advocated by Anderson et al.~\cite{anderson2005overcoming}, \archname begins by directly confronting one highly plausible source of stagnation, namely the specific virtualization that is \textit{inherent} to the Internet architecture.\fi
The guiding idea of \archname is to overcome the Internet impasse {\em by going under it}, i.e., by putting the EBA spanning layer beneath the specific virtualization that is \textit{inherent} to the Internet architecture.
In the Internet stack below the Network Layer, the Link layer models services that are local to network nodes, or that connect them in local area networks. Aggregating these low level resources in the implementation of IP to create wide area services has two different purposes:
\begin{itemize}
\item It {\em virtualizes} the variety of local services, enabling interoperability through the adoption of a common model.\footnote{In this context the meaning of {\em virtualize} is "to transform into a computer-generated version of itself which functions as if it were real", not necessarily the related meaning of "creating multiple computer-generated versions of itself".}

\item It {\em hides} the complex and dynamic topology and behavior of local infrastructure by providing an abstraction that restricts client visibility into local resources.
\end{itemize}
The premise of \archname is that we can separate these two tasks. First we define a \textit{local} virtualization of lower layer services and resources (including storage and processing) to enable interoperability.
Then this local virtualization enables heterogeneity of higher level services, to be defined at the Network layer. This preserves interoperability of implementation while avoiding the ``ossification'' that comes from imposition of a uniform abstraction at the Network layer; {\em global forwarding becomes a choice, not an intrinsic necessity}.

\begin{figure}[hbt] \centering
	\includegraphics[width=3in]{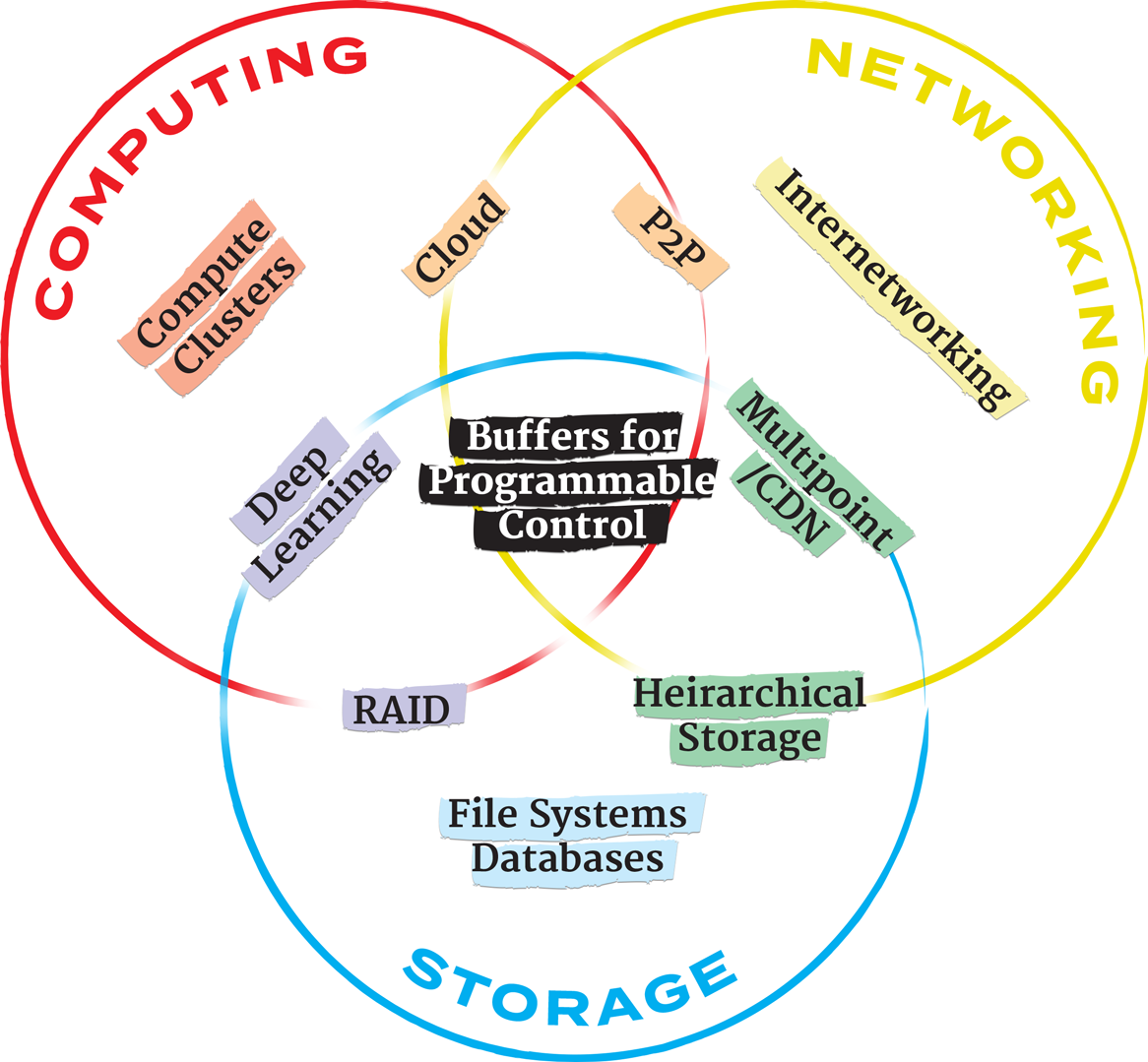}
	\caption{Buffers, i.e. mechanisms for data persistence, are used to implement abstractions across the storage, networking, and computing stovepipes. The common service layer of the EBP platform exposes a generic interface to buffers. \label{Venn}} 
\end{figure}

As its name suggests, the design of the Local layer virtualization on which \archname builds focuses on the \textit{buffer}, a data persistence resource of limited size that captures values generated by adjacent devices within a system and holds them until they can be delivered to others (perhaps with very low latency).
Buffers are the fundamental building block of all sequential digital circuits. Familiar higher level abstractions, such as data storage, movement and computation, are implemented using them (Figure~\ref{Venn}).
For example, if we analyze the implementation of the process abstraction, we find that it consists of the management of state stored in buffers of different size and characteristics (e.g., registers and pages).
From this point of view, packets delivered end-to-end and processes that transform data on a single node are closely related, and can be unified by expressing them both in terms of a generic buffer abstraction.
The buffer abstraction also applies to storage at the block or object level, completing the three-way convergence.

Thus \archname can account for all of the resources of the network node, including the storage and processing resources (normally excluded in representations of the ``network stack'') which are used for system configuration and the calculation of forwarding routes. The hypothesis on which \archname is based is that higher level operations can be implemented with the greatest level of flexibility and efficiency by expressing them in terms of primitive buffer operations and then aggregating those operations explicitly to create higher level services. 
\archname exposes the storage and computational resources that are now used to implement datagram delivery, but which are hived off from other mechanisms and policies by the adoption of the Internet spanning layer.
In doing so, \archname defines a converged infrastructure model in which transfer, persistence and processing functions can be combined freely to enable a wide variety of programmable and stateful services and protocols.

The alternative perspective on convergence is the current siloed approach: construct higher level abstractions that encapsulate their constituent buffers and operations on them and then compose those abstractions in overlays or middleboxes (either physical or virtual). 
%By enabling network layer datagrams to be inspected and modified by services at intermediate nodes. 
For instance Network Function Virtualizaiton (NFV) seeks to provide routing configurability, programmability and stateful services by marrying end-to-end datagram delivery with some form of process management~\cite{chiosi2012network}.

\archname is a much more direct approach which uses the fact that datagrams are not physical entities but abstractions (distributed processes) that aggregate local operations on local buffers along a path.
Processes running on intermediate nodes are also abstractions of operations performed on memory pages implemented through instruction execution.
Rather than seeking to combine ICT silos through the creation of a high level overlay, 
\archname enables the implementation of services  from a small number of general and interoperable of buffer operations.
Defining an infrastructure in terms of buffer operations rather than higher level silos of storage, networking and computation provides a more direct path to convergence of those silos.

\section{Exposed Buffer Architecture}
While \archname  posits that all computer systems can be expressed as the aggregation of operations on a generic abstraction of persistent data buffers, such buffers have many implementation technologies, come in different sizes, are connected to other buffers in different ways, and have a variety of transformative operations defined on them. The smallest buffers typically considered in the design of large scale systems are single bit cells, and the largest are contiguous storage extents consisting of gigabytes.
Connectivity ranges from the inputs and output paths of FIFO buffers to highly interconnected local area networks.
Storage duration might be measured in microseconds or in years.
Transformation operations range from bitwise Boolean functions to complex pattern recognition.

Importantly, to maximize deployment scalability, the \principlename principle tells us that the \archname buffer abstraction must represent the aggregation of information about fundamental device characteristics, but without assuming characteristics that are not strongly enabled by the underlying physical devices. The \archname buffer service conforms to this design criterion:

\begin{itemize}
\item No persistent medium is unbounded, so EBP buffers can be limited in size.
\item No persistent medium is permanent, so EBP buffers can be limited in duration.
\item No persistent medium has perfect fidelity, so EBP buffers are best-effort.
\item No buffer operation has unbounded inputs or outputs
\item No buffer operation can guarantee correct completion
\item An operation acts on buffers that are physically colocated or locally connected

\end{itemize}

But despite these weak assumptions, all systems can be described in terms of a uniform abstraction of such buffers and operations on them, and many can be implemented using interoperable mechanisms, (subject to performance constraints).  The set of primitive functions required is small:
\begin{itemize}
\item allocation of a buffer,
\item storage and retrieval of data in a buffer over time,
\item transfer of data between locally connected buffers, and
\item transformation of data in a set of buffers that have some common locality
\end{itemize}

From this point of view, a datagram is characterized as the transfer of data from an originating buffer, through a series of buffers either internal to network nodes or connecting buffers in adjacent nodes, to eventually reach a destination buffer.
Forwarding is a local operation that connects input and output buffers, or connects those buffers to the memory pages of the router. Computation and processing, on the other hand, can be characterized as the local transformation of a set of buffers that implement the pages of a process address space, including the stored program.
Some of those buffers may reside in an operating system, and may be shared in the implementation of multiple concurrently active processes.
Other buffers are special purpose  registers or memories (e.g., caches) connected directly to fixed datapaths and control functions.
Such specialized and optimized implementation interferes with interoperability, but these elements are still recognizable as buffers (see Figure~\ref{Venn}).

Conventional implementation strategies distinguish datagrams in flight from the pages of an executing process in ways that then require complex workarounds to allow them to be combined.
%
%In Software Defined Networking a routing table is constructed that requires deeper packet inspection and more complex processing to determine forwarding decisions.
%This very slightly allows transformation of data held in buffers to be applied to the datagram as it resides temporarily in the data plane of an intermediate node.
%
In "middleboxes" or Network Function Virtualization, a datapath is created from the datagram forwarding path to the address space of processes implemented using timeslices of a conventional ISA. 
These timeslices implement complex compound operations on the state of the process and the datagram.
%In the NFV model, the aggregation of those timeslices implements a virtualization of the network node.

%SDN achieves a high level of performance by greatly restricting the generality of the data path and the in-transit processing.
NFV implements a restrictive datapath in order to maintain the performance of one function, datagram forwarding, while enabling the possibility of additional network processing.
Exposed Buffer Processing (EBP) is an approach that first maximizes generality and interoperability and only then considers how  performance can be achieved, either through the use of optimizations that leverage converged network resources (data logistics) or by optimization and possibly acceleration along the fast datagram forwarding path.

\section{Exposed Buffer Processing}
Exposed Buffer Architecture is a general approach to the design of systems, and applies to many implementation strategies at many scales. 
For example, one interpretation of RISC architecture is that in decomposing multi-cycle operations into micro-operations connected by explicit register operations, it is an application to datapaths within a processor of the same approach suggested by Exposed Buffer Architecture.
On a different scale, the use of replicated servers in Content Delivery Networks accompanied by topology-aware DNS resolution is also an application of Exposed Buffer Architecture, albeit one that is then hidden from end users through spoofing of the URL display in browsers.

\begin{figure}[hbt] \centering
	\includegraphics[width=3in]{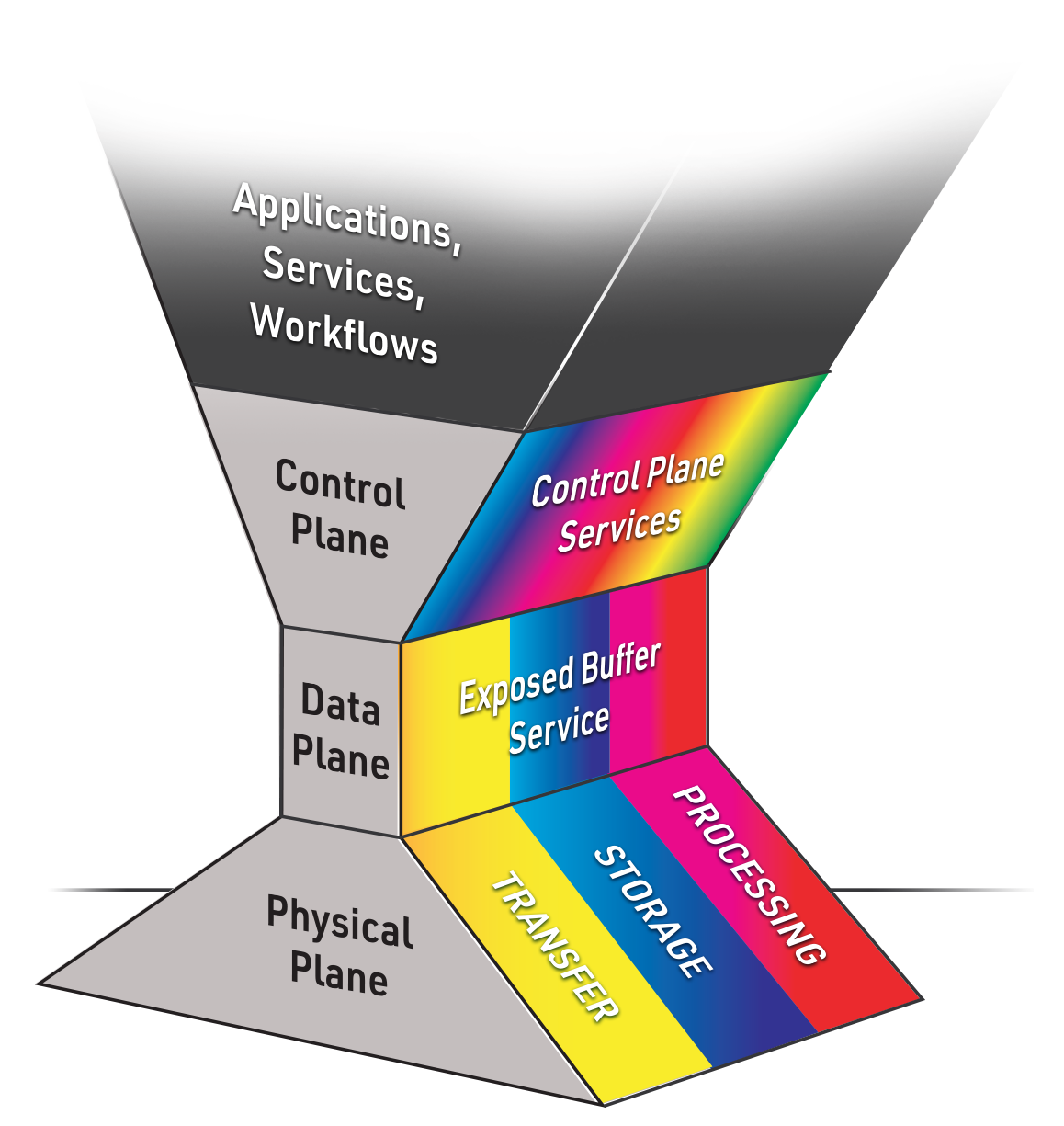}
	\caption{The EBP Layered System "Anvil". \label{anvil}} 
\end{figure}

Exposed Buffer Architecture~\cite{beck2019interoperable} adopts a model in which resources that implement persistence and transformation of buffers local or adjacent to a node are recognized as resources of the physical layer.
The layer above the physical layer (a generalized ''local layer'') implements a virtualization of those general resources through a minimal API or local area protocol. As described above, the EBP service defines a means of allocating, transferring and transforming data buffers using resources local to the node or within a LAN. That can be thought of as a passive execution mechanism similar to the datapath of a processor, but extending to transfer among adjacent nodes. We call the infrastructure that implements this primitive EBP service, which is a generalization of the Link Layer of the Internet stack, the {\em data plane}. In the Exposed Buffer Architecture, all services with greater levels of functionality are composed by aggregating the primitive EBP service in the the \emph{control plane}, which is a generalization of the Internet's Network Layer (Figure~\ref{anvil}).

\subsection{Mapping Traditional ICT Silos to \archname}

Within Exposed Buffer Architecture, a datagram is modeled as a set of buffers that are allocated along a path of nodes connected by local area hops.
Forwarding is modeled as a transformation of the datagram header that results in a forwarding choice which is then the basis of an update transformation to the header and a buffer transfer across a single hop.
The buffers along the path may be allocated dynamically (FIFO buffering is interpreted as a highly transient form of buffer allocation) or persistently (as in reservation for the purpose of QoS).

Within Exposed Buffer Architecture, a process is modeled as a set of buffers (pages) that can be transformed by computation on the nodes where they are allocated.
Transformations may represent operations within inputs and outputs (potentially functional) that are executed in dataflow fashion or following order specified by program dependences.
If a process is expressed as a stored program augmented with operating system services (e.g., a POSIX process) then the pages of the program are modeled as buffers and the operating system state are additional buffers that are accessible only through the intervention of the next higher level in the stack,
which we call the {\em control plane}.
Note that in this terminology, the Network Layer of the Internet stack is one possible service that can be implemented in the {\em control plane}.

Within Exposed Buffer Architecture, a file is modeled as a set of buffers (storage objects) that are allocated from within volumes located at nodes where data is placed.
Data encoding for error detection and correction and redundancy for performance are implemented through the control plane.
Fault tolerance and performance enhancing algorithms as well as data migration and placement according to other policies are similarly not implemented by the buffer service.
However, specific operations such as calculating hashes or reconstructing data when storage corruption occurs may be implemented within the buffer service using  buffer transformations.

\subsection{Security in Exposed Buffer Networks}
The need to provide secure wide area service under the Internet model has been a long-standing problem with no obvious solution in sight.
One reason for this lies in the choice of end-to-end datagram delivery as the universal service that defines the common ground of the Internet.
If we consider routers to control access to all the resources of the Internet that are shared in the wide area, then the means by which (subnets consisting of) routers share those resources is through the exchange of routes (e.g., through BGP).
Once a route has been advertised to a peer, a router has little basis on which to make admission decisions, and transit through the router is provided to all.
Transit through a router implies that it will allocate whatever resources it has access to from the next router on the path, in effect acting as a full proxy for the sender.

In contrast the local buffers and operations on them comprise the resources of an Exposed Buffer Architecture.
Securing the resources of a local node is an easier task because the mechanisms used do not need to scale to the entire community of network users, only to those with direct access to the node.
In Exposed Buffer Architecture, access to the resources of non-local nodes is done through services implemented in the control plane, through service-specific peering mechanisms and agreements that can be specialized to the community served.
Thus, global routing need not be supported by any service on a particular node, if it chooses not to participate in the kind of unmediated communication service that defines the Internet.

\subsection{EBP in Overlay: The Internet Backplane Protocol}

Exposed Buffer Processing is the name for an (as yet undefined) implementation of the architecture within the current Internet stack, although not adhering to all of its original architectural principles (in the same way that Content Delivery Networks do not).
The current implementation of Exposed Buffer Processing is the Internet Backplane Protocol, which is a full overlay implementation that {\em does} adhere to Internet principles (specifically, isolating overlay routing from Link Layer topology information and mechanisms)~\cite{Beck02anend-to-end}.
As we discuss in Section~\ref{sec_implementations}, there is an evolutionary path from IBP to more native forms of EBP that are encapsulated within IPv6 datagram delivery or are implemented directly on the local layer (a so-called ``clean-slate'').

IBP is implemented as a set of RPC-over-TCP calls, and so it does not have the inherently local characteristics of EBP (although these can be imposed using firewalls or SDN routing).
The core functionality of EBP is implemented in a few simple and general calls:
\begin{itemize}
\item {\bf Allocate} a buffer on a specific node, specifying the size and duration explicitly.
The client receives capabilities (random keys) that are the only names by which the allocation can be referenced and which also implement per-allocation access control.

\begin{figure*}[hbt] \centering
	\includegraphics[width=1.0\textwidth]{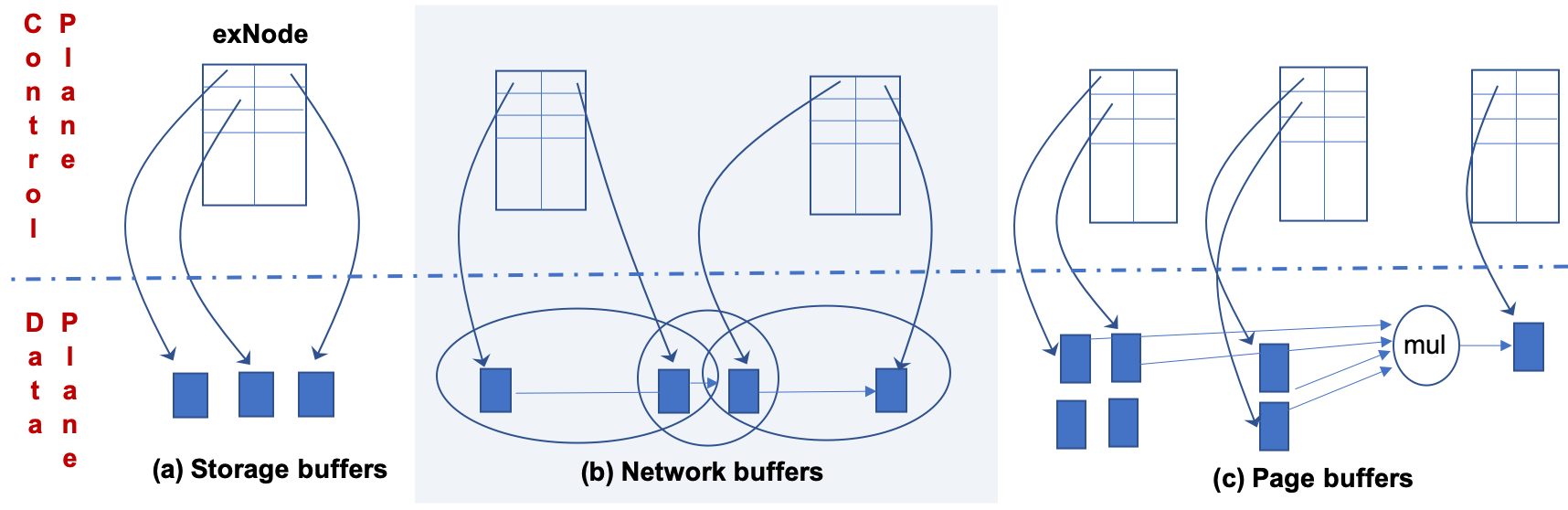}
	\caption{Structural metadata in the \archname control plane uses a common interface to buffers in the \archname data plane to manage diverse operations. \label{2planes}} 

\end{figure*}

\item {\bf Write} data to a buffer or {\bf read} data from a buffer.
This allows clients which are applications to implement higher level operations directly.

\item  {\bf Transfer} of data between two buffers on one data plane node or between two by specifying the keys of both sender and receiver. These data plane nodes must be adjacent (reachable through Internet routing).

\item {\bf Transform} data stored in a set of buffers on a single data plane node by invoking the execution of an operation on that node.
Such operations are bound to a global namespace and must be implemented by the node through some mechanism other than invocation.
This means that operations do not necessarily imply the use of on-demand or mobile code mechanisms.
\end{itemize}

All of these calls and the services they implement are best-effort, meaning that they make no guarantees of completion, integrity or correctness either immediately or over time.
All resource allocations, including buffer space, transfer bandwidth, and computational resources, can be capped by the operator of the data plane node (in analogy to the network Maximum Transfer Unit).
While data plane nodes may implement strong services and provide guarantees, all knowledge of and control over QoS is implemented by control plane services.

\section{Structural Metadata and the Control Plane}
There are applications that can make direct use of the resources of the data plane
to allocation, move and transform data.
However, those functions are quite primitive and local, and in general must be aggregated to implement abstractions that are more easily usable by application developers.
IBP clients can also implement control plane services which they then offer to applications or to other services.

The analogy in Networking is the aggregation of IP best effort datagram delivery into connected TCP streams.
In computation, it is the aggregation of primitive operations or VM time slices into orchestrated collections of threads.
In storage, it is the aggregation of disk blocks or storage objects into file systems or databases.

The primitive resources of the data plane are named and controlled using capabilities, which act as pointers that are either local to a node or in the case of transfer, dereferenced across a LAN  adjacency.
These capabilities can be organized into data structures, along with additional control-oriented information for the purpose of implementing higher level abstractions and functions.
We call these data structures {\it structural metadata}.

In Networking structural metadata is found in packet headers and connection state.
In computation, it is the process descriptor, memory management and other control registers.
In storage, it is the inode, B-tree or other storage data structure used to organize and access data blocks. (Figure~\ref{2planes})

%\subsection{The exNode}
In \emph{Logistical Networking}, the data structure created to organize structural metadata composed of IBP capabilities is the \emph{exNode}~\cite{Bassi01MobileManagement}.
Modeled after the function of the Unix file system's inode data structure, the exNode aggregates multiple IBP allocations to implement a large contiguous data extent addressed linearly.
It also aggregates redundant storage allocations to implement RAID-like fault tolerance either locally or using storage distributed across the wide area network.
%\subsection{Logistical Networking}

The role of structural metadata is just as important in Networking as in Storage.
In the conventional Internet stack, such metadata is distributed in packet headers as well as maintained in TCP stream state at endpoints.
In Logistical Networking, we can use the exNode as a representation of stored data and more transient data structures within the control plane to represent and control data as it moves through the network.
Because of the separation between the control and data planes in EBP architecture, structural metadata may remain physically centralized even as it represents data that is distributed and mobile within the data plane.

%\subsection{Converged Computation}
The structural metadata associated with a computational process is typically held by an operating system in the form of memory management data structures (page tables), task descriptors, hardware contexts (the contents of registers) and other miscellaneous data structures.
Typically, these are expressed in terms of physical addresses and other highly local information that ties each task to one node or to a shared-memory environment.

When data being computed on takes the form of file-like extents, the exNode can take on the role of a page table (much as in memory-mapped I/O).
However as with networking, data sometimes has to be managed or moved during computation in ways that may be better represented by other data structures.
In cases where computation is applied to stored data over long periods of time, it may be useful to generalize the exNode to represent computationally oriented data management.
Examples are the localized expansion of data (e.g. decompressed, unpacked or transposed) for faster access or transformation).

\subsection{Distributing the Control Plane}
Separating structural metadata from "payload" data has a number of advantages.
The metadata is typically much smaller (less than  1\% of the data data size, depending on representation) and so can be managed  flexibly and efficiently.
To the extent that EBP is implemented using a local (or in the case of IBP, wide area) networking environment, a centralized control plane process can manage data stored in a large number of individual data plane nodes.
This can expedite implementation of the control plane, although the resulting systems are limited to environments in which the data plane is sufficiently well connected to the more centralized control plane to enable such separation.

A further stage in the implementation of the control plane is to enable the distribution of control functions.
While this complicates the implementation of the control plane, it also generalizes its functionality and enables deployment in a wider variety of network environments.
A distributed control plane is more scalable, facilitating services in which   control functions use more substantial resources.
Even more important, in environments where communication between local area domains (subnets) needs to be restricted for security or policy reasons, peering between those domains can occur through coordination between control and data plane nodes at the borders.
Thus control plane functions can be implemented in a more centralized or more distributed manner according to specific performance and policy requirements.

A further evolution of the implementation of the control plane is using data plane resources to store metadata and to perform transformations.
An example would be storing exNodes in data plane storage allocations, represented by a higher level exNode.
That mode of storage could be supported by data plane operations which search through the exNode data structure and extract metadata.
The categories of control plane functions that might then be supported using data plane resources could expand to enable the control plane to remain less resource-intensive and thus amenable to a more compact implementation.

\section{Application Design for the Digital Continuum}

Application design in a distributed environment has many challenges, including communication and application partitioning. These two are related, because the characteristics of communication (bandwidth, latency, reliability, security) determines the design space of application partitioning. Typically, communication mechanisms and abstractions are designed to meet a general notion of application requirements. Interface and algorithm designers must then work within these constraints.

One direct approach that is often taken to distributed application design is to create an {\em articulated} system, in which the application is specified and implemented as a set of components connected by well-defined interfaces. This approach is popular in cases where some components are implemented in a semi-static manner and may have to be shared between different distributed applications. The reasons for these constraints vary but can include those components being implemented on small devices in remote environments. We find these characteristics in edge computing environments.

A classic example of an articulated approach to distributed applications is the Java applet deployed to implement a widget operating within an HTML page. The applet deployment and invocation mechanism is defined within the HTML standard and implemented by Web browsers. Some applets communicate with the servers on behalf of which they are deployed, using protocols that may themselves be published or standardized.
This enables the use of the applets in a variety of applications. Such sharing of applet implementations reduces the complexity of distributed application development and can also result in improved deployment performance through the caching of applet code by the browser.

In contrast to the articulated approach, a {\em continuous} approach defines an execution model that is as uniform as possible, describing the application in terms that are more independent of the resource topology into which it will be mapped and specifying communication between components more abstractly. In the extreme, a continuous approach is completely independent of the resource topology on which it will be deployed, specifying computing in terms of more abstract intentions. 
By not explicitly mapping the application onto the resource topology, the continuous approach requires static tools such as a compiler or dynamic mechanisms such as a runtime scheduler to determine and implement the  mapping. 
This creates a tension between the degree of continuity in the application architecture and the performance achieved through the use of such tools.

In the case of edge functionality, a continuous approach could model the state of the server-resident application, the contents of network buffers and content caches, the state of the edge computation and the interaction with the human interfaces all as buffers.
Computation, storage and processing would all be expressed in terms of operations on such buffers, including movement between them over the wide area.
Interposition of middlebox functionality could then also be modeled as the introduction of buffers within the network topology.
Effective management of such a continuous buffer model requires the application of many types of metadata by control plane mechanisms, including some that may be manually configured, obtained from dynamic network services, or measured by the application itself.
While this highly layered approach places a burden on the designers of the application software stack, it also enables a huge degree of design flexibility without the (sometimes insurmountable) overhead of changing articulation points and redefining widely accepted interfaces.
More examples have been described in the literature~\cite{beck:2020:MASS}.

The optimal point in the design of application architectures is not fixed but varies according to the nature and complexity of the applications and the partitioning and scheduling challenges of the execution environment. To further complicate matters, these characteristics change over time as new tools are developed as the environment itself evolves. Given these factors, one way to address the design challenge is to define an execution model that captures the greatest degree of commonality among the possible compositions while enabling the greatest variety of specific choices.

The key to such generality is to expose the most fundamental components out of which distributed applications are constructed, in a form that is maximally weak, simple and general~\cite{beck2019hourglass}. The simplest and most common model of the various storage, networking and computational elements that comprise the specification and implementation of a distributed application is provided by Exposed Buffer Architecture.

\section{Implementation and Validation}

\subsection{EBP Implementation}
\label{sec_implementations}
Deployment of network architecture innovations is complicated by the dominance of the Internet Protocol Suite at the Network and Transport layers and the need for stability in both commodity and research networks that are critical important day-to-day operations.

Overlay solutions that are compatible with the architectural rules of the current network inhibit effective use of information and mechanisms that are restricted to the lower layers (e.g., routing and security policy).
Native solutions can be deployed when access to the lower layers is possible, for example when the innovative service is being deployed by the owner of the physical infrastructure (e.g., Content Delivery and Cloud networks) or by their business partners.
Extensions of commodity networking environments (e.g., NFV) must remain compatible with end user interfaces while allowing greater configurability or programmability of network nodes. 
These approaches have not yet yielded infrastructure that exhibit a high degree of deployment scalability.

We propose an approach to deploying EBP that proceeds using a combination of overlay and native mechanisms.
An EBP protocol can be encapsulated within IP packets, enabling it to share physical infrastructure with commodity Internet traffic, but virtual LANs or SDN can be used to keep EBP traffic from being routed between networks.
In essence, IP is thus used as a convenient local area networking solution.
To connect local EBP domains Internet connectivity between peering control plane nodes can be used.
This approach enables dedicated control plane nodes to be deployed, but the EBP control plane can also be implemented within the container execution capability of NFV-enabled routers.

\subsection{\archname Validation}
\label{validation}

Validation of the design of wide area infrastructure is a very challenging topic.
This is due in part to the goal of deployment scalability, which is defined in terms of the ultimate acceptance of the design and its wide utility in application domains not necessarily anticipated in advance.
It is also due to the fact that building wide area information infrastructure tends to be a highly exclusive undertaking.
While successive epochs may overlap, there is generally a single dominant model at any given time.
During the period when telephone and radio frequency broadcast media were dominant, the idea that packet networking would arise as a force for convergence that would ultimately dominate was widely unthinkable.
In the current period of packet networking and Cloud data centers, the idea that further convergence could arise from the convergence of storage, networking, and computation silos may seem similarly implausible to many.

Various aspects of the proposed Exposed Buffer Processing model have been validated in a number of different ways.
As described above, the Internet Backplane Protocol and Logistical Networking are overlay implementations of Exposed Buffer Architecture which have been successfully used to build a wide variety of different services and to display experimental performance in overlay, as described in the literature~\cite{beck2019interoperable,beck2019data}.
However, that record of success was achieved over the past 15 years, so the implementation details and performance levels do not match the current vision of EBP as a unifying mechanism at a much more fundamental level.
It may seem a lot to ask of a field that advances so rapidly in implementation technology to extrapolate the lessons of past decades to design for the future.
Reestablishing the fundamental validity of the model experimentally would take a development effort comparable to the original overlay implementation-- millions of dollars applied in a multi-year project.
Technological tradeoffs do not occur on a timetable, and design insights gained in a previous era may remain relevant even as times change.

Another mode of validation is the use of formal reasoning in the design process.
A search for the formal basis for  End-to-End Arguments~\cite{Saltzer:1984:EAS:357401.357402} led to the formulation of the Deployment Scalability Tradeoff, the core of which is expressed in the Hourglass Theorem~\cite{beck2019hourglass}.
Expressed in the language of Program Logic, this theorem provides a theoretical basis for the intuition of operating system and network designers that the "weakest" common interface should be used to support a required set of applications in order to also maximizing deployment scalability.
Such an abstract foundation may seem inaccessible to some infrastructure architects, but it is supported by many historical examples taken from as the design of the Unix operating system and the Internet stack. 
Of course, there may be doubts about whether abstract principles that help explain the historical evolution of successful and unsuccessful infrastructures are applicable as validation of a new paradigm at a different layer of the infrastructure service stack.

Ultimately, the question for the skeptical reader is whether the radical approach to convergence that we suggest is plausible.
We do not ask readers to conclude that we have the ultimate protocol or implementation of the principles involved, but only entertain the idea that following this design approach has the potential to lead to a new epoch in innovation and interoperability.
The question is not whether the efficacy of this architecture has been proved, but whether investigation and development in this direction deserve the attention of a community that has for decades been frustrated by the constraints imposed by their own early successes.

\section{Related Work}
Exposed Buffer Architecture grew out of responses to the limitations of stateless networking.
Efforts to address application demands which are not well addressed by the conventional client/server paradigm have led to a sequence of solutions from FTP mirroring to Web caching, middle boxes and network-embedded virtual machines running on multi-tenant infrastructures: PlanetLab, GENI Racks, and most recently NFV.
At the same time, the demand for utility storage and computing services have given rise to a sequence of solutions based on distributed data centers from early online services (e.g., Compuserve and AOL) to the computational Grid and most recently the National Research Platform.
Edge devices and Fog data centers serving the Internet of Things combine the functionality typically found in clouds with widespread deployment in embedded environments.
One increasingly important category of services that might naturally be expected to be localized that currently are implemented as cloud services supported by myriad proprietary edge devices are home entertainment, security and management applications.
Exposed Buffer Processing uses the resources of network nodes to solve problems of data logistics (combined movement, storage and computing) in a common, primitive, interoperable form rather than through the high-level services of silos. 
%In that sense, EBP proposes a reconsideration of design choices that are considered foundational to those silos, allowing for the construction of services that blur conventional boundaries to enable flexible and efficient solutions to problems of data logistics.

\section{Conclusion}
In this paper we have presented Exposed Buffer Architecture, which can enable a radically expanded class of widely deployable network services implemented at the Network layer.
\archname proposes to move the virtualization layer that provides interoperability/portability in the implementation of applications to the local layer, which is a generalization of the Internet's Link layert to include storage and processing resources.
This approach enables backward compatibility with current Internet Protocol clients in Network layer services, but does not require it.
It defines a converged platform for implementing new services as alternatives to the Internet.

\bibliographystyle{IEEEtran}
\bibliography{ebpbib.bib,ebabib.bib,hglass.bib,expeditions.bib}

\end{document}